\documentclass[hidelinks]{article}

\usepackage{arxiv}

\usepackage[utf8]{inputenc} 
\usepackage[T1]{fontenc}    

\usepackage{amsmath, amssymb, amsthm, mathtools, mathrsfs} 

\usepackage{hyperref}       
\usepackage{url}            
\usepackage{amsfonts}       
\usepackage{nicefrac}       
\usepackage{microtype}      
\usepackage{graphicx}
\usepackage{doi}
\usepackage{natbib} 
\usepackage{pdflscape} 
\usepackage{acro} 
\usepackage{bookmark} 
\usepackage{nameref} 

\usepackage{tabularx}
\usepackage{booktabs}
\usepackage{caption}
\usepackage{tabu}
\usepackage{rotating} 
\usepackage{makecell} 
\usepackage{multirow} 
\usepackage{float} 
\usepackage{array}
\usepackage{arydshln}
\newcolumntype{P}[1]{>{\centering\arraybackslash}p{#1}}
\newcolumntype{R}[1]{>{\raggedleft\arraybackslash}p{#1}}

\usepackage{chngcntr}
\counterwithin{figure}{section}
\counterwithin{table}{section}
\counterwithin{equation}{section}

\usepackage[dvipsnames]{xcolor} 

\usepackage{algorithm}
\usepackage{algpseudocode}

\usepackage{bm} 

\title{Overcoming the Representational Limitations of Amplitude Encoding for Quantum Neural Networks}

\author{
    \href{https://orcid.org/0009-0002-1502-3670}
    {\includegraphics[scale=0.06]{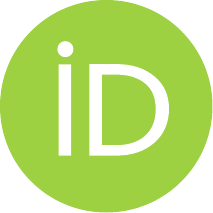}
    Vincent Gurgul} \\
	Chair of Information Systems\\
	Humboldt-Universität zu Berlin\\
	Unter den Linden 6, 10117 Berlin, Germany\\
    vincent.gurgul[at]hu-berlin.de\\
	\And
	\href{https://orcid.org/0000-0001-7089-8870}
    {\includegraphics[scale=0.06]{orcid.pdf}
	David Prinz} \\
	Lamarr Institute for Machine Learning and Artificial Intelligence\\
	Rheinische Friedrich-Wilhelms-Universität Bonn\\
    Friedrich-Hirzebruch-Allee 6, 53115 Bonn, Germany\\
    prinz[at]bit.uni-bonn.de\\
}

\renewcommand{\headeright}{}
\renewcommand{\shorttitle}{Overcoming the Representational Limitations of Amplitude Encoding for Quantum Neural Networks}

\hypersetup{
pdftitle={Overcoming the Representational Limitations of Amplitude Encoding for Quantum Neural Networks},
pdfsubject={quantum, computing},
pdfauthor={Gurgul, V., Prinz, D.},
pdfkeywords={quantum machine learning, quantum neural networks, variational quantum circuits, universal approximation theorem, classical data encoding},
}

\DeclareAcronym{VQC}{
	short = VQC,
	long = Variational Quantum Circuit
}
\DeclareAcronym{DDPG}{
	short = DDPG,
	long = Deep Deterministic Policy Gradient
}
\DeclareAcronym{QPU}{
	short = QPU,
	long = Quantum Processing Unit
}
\DeclareAcronym{PQC}{
	short = PQC,
	long = Parameterized Quantum Circuit
}
\DeclareAcronym{QNN}{
	short = QNN,
	long = Quantum Neural Network
}
\DeclareAcronym{QUBO}{
	short = QUBO,
	long = Quadratic Unconstrained Binary Optimization
}
\DeclareAcronym{QAOA}{
	short = QAOA,
	long = Quantum Approximate Optimization Algorithm
}
\DeclareAcronym{VQE}{
	short = VQE,
	long = Variational Quantum Eigensolver
}
\DeclareAcronym{NISQ}{
	short = NISQ,
	long = Noisy Intermediate-Scale Quantum
}
\DeclareAcronym{CQM}{
	short = CQM,
	long = Constrained Quadratic Model
}
\DeclareAcronym{QA}{
	short = QA,
	long = Quantum Annealing
}
\DeclareAcronym{PCA}{
	short = PCA,
	long = Principal Component Analysis
}
\DeclareAcronym{AUC}{
	short = AUC,
	long = Area Under the Receiver Operating Characteristic Curve
}
\DeclareAcronym{AUPR}{
	short = AUPR,
	long = Area Under the Precision--Recall Curve
}
\DeclareAcronym{OvR}{
	short = OvR,
	long = One-vs-Rest
}

\begin{document}
\maketitle

\begin{abstract}
Quantum Neural Networks (QNNs) offer a framework for learning functions from classical data, but their expressive power depends critically on how inputs are encoded into quantum states. Angle encoding can require resources that scale with the number of features, whereas amplitude encoding represents $d$ features using only $\mathcal{O}(\log_2 d)$ qubits. This compactness comes at a representational cost: normalization makes conventional amplitude encoding invariant to nonzero real rescaling of the input, up to an unobservable global phase. We show that this invariance induces an exact, architecture-independent restriction on the resulting hypothesis class. No downstream quantum circuit can distinguish inputs differing by nonzero real scaling. We develop a general condition under which deterministic preprocessing can remove this ambiguity and show that nonlinearity alone is insufficient. Based on this analysis, we introduce a fixed nonlinear feature map that provably distinguishes all real inputs under subsequent amplitude encoding, without adding trainable parameters and while retaining logarithmic qubit scaling. We evaluate the resulting QNN on three canonical classification benchmarks and a nonlinear synthetic dataset against conventional amplitude and angle encoding, as well as a classical neural network. Across these tasks, the proposed encoding consistently improves classification accuracy over both quantum encoding baselines by 21--44 \%. Our results establish normalization-induced scale invariance as a concrete representational bottleneck of amplitude encoding and demonstrate that it can be removed while preserving a compact quantum representation, marking a step toward quantum models capable of capturing arbitrary relationships in classical data.
\end{abstract}



\section{Introduction}

Recent advances in gate-based quantum computing have enabled \acp{VQC} as \acp{QNN} for supervised learning. To process classical data, these models must first encode an input vector into a quantum state \citep{ciliberto_quantum_2018}. This encoding determines both the relationships in the data that the circuit can represent and hence the expressivity of the resulting model \citep{schuld_effect_2021-1}. Furthermore, it defines how resource requirements scale with input size, making data encoding a major bottleneck in the pursuit of quantum advantage in machine learning \citep{cerezo_challenges_2022}.

Angle encoding maps features to quantum-gate parameters and can require resources that scale with the number of features.
Amplitude encoding instead represents a \(d\)-dimensional input using only \(\mathcal{O}(\log_2 d)\) qubits.
However, its required normalization introduces a fundamental ambiguity. For any nonzero real input \(\mathbf{x}\) and nonzero scalar \(\lambda\), amplitude encoding maps \(\mathbf{x}\) and \(\lambda\mathbf{x}\) to states that differ at most by a global phase and are therefore indistinguishable.
Thus, no downstream quantum circuit, irrespective of its architecture or trainable parameters, can distinguish inputs related by nonzero real scaling \citep{havlicek_supervised_2019, mottonen_transformation_2004}.

This work investigates this normalization-induced scale invariance as a representational limitation of amplitude-encoded QNNs.
We derive a general condition under which deterministic preprocessing removes the ambiguity and show that nonlinearity alone is insufficient.
Based on this analysis, we introduce a fixed nonlinear feature map that distinguishes all real inputs after amplitude encoding, without trainable parameters and while preserving logarithmic qubit scaling.
We evaluate the resulting variational quantum classifiers on three canonical benchmarks and a nonlinear synthetic dataset, comparing them with conventional amplitude encoding, angle encoding, and a classical neural network.
The results show that recovering scale information can substantially improve predictive performance while retaining the principal resource advantage of amplitude encoding.


\section{Variational Quantum Circuits as Quantum Neural Networks}

\acp{PQC} apply a unitary transformation \(\mathbf{U}(\boldsymbol{\theta})\) with tunable parameters \(\boldsymbol{\theta}\). \acp{VQC} are PQCs whose parameters are optimized with respect to a cost function derived from measurement outcomes \citep{benedetti_parameterized_2019}.
In this work, the term \ac{QNN} refers specifically to VQCs, although it is also sometimes used for other quantum learning models \citep{schuld_quest_2014}.

\vspace{-5pt}

\begin{figure}[H]
    \centering
    \includegraphics[width=0.56\textwidth]{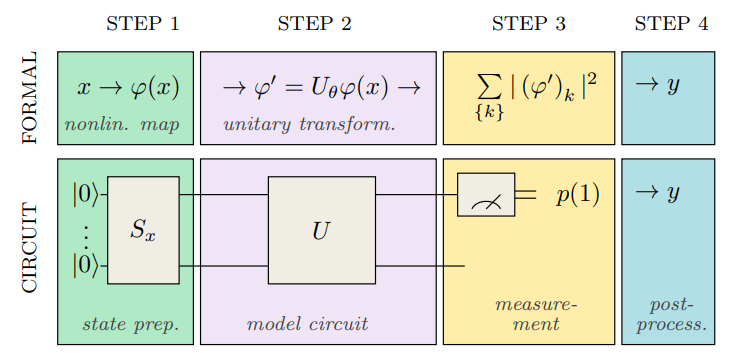}
    \vspace{8pt}
    \caption{Illustration of the four steps involved in inference with a \ac{VQC} \citep[Adapted from][]{schuld_circuit-centric_2020-1}.}
    \label{fig:vqc}
\end{figure}

\vspace{-10pt}

As shown in Fig.\ \ref{fig:vqc}, inference with a VQC consists of four steps: (i) encoding classical data \(\mathbf{x}\) into a quantum state, (ii) applying a parameterized transformation \(\mathbf{U}(\boldsymbol{\theta})\), (iii) measuring observables, and (iv) post-processing the outcomes to produce predictions or evaluate a cost function.
The trainable circuit typically alternates single-qubit rotations, such as \(\mathbf{R}_X(\theta)\), \(\mathbf{R}_Y(\theta)\), and \(\mathbf{R}_Z(\theta)\), with entangling gates such as CNOTs.
Measurements are probabilistic, so expectation values must be estimated from repeated circuit executions (shots).
These estimates define a real-valued cost function \(C(\boldsymbol{\theta})\), which is minimized by a classical optimizer.
Gradients can be evaluated analytically for many parameterized gates using the parameter-shift rule \citep{mitarai_quantum_2018,schuld_evaluating_2019}.
With suitable encodings and circuit architectures, \acp{VQC} can approximate broad classes of functions similarly to classical neural networks \citep{schuld_effect_2021}.


\section{Problem Statement and Methodology}

In order to be processed by a \ac{VQC}, classical data must be encoded onto a quantum state. There are multiple strategies for this encoding, and the choice of method strongly influences both the expressive capacity of the circuit and its resource requirements.

\subsection{Existing Classical Data Encoding Methods and Their Shortcomings}

The fundamental classical data encoding approaches are basis encoding, angle encoding, and amplitude encoding \citep{schuld_effect_2021}. Additionally, there exist derivatives of angle encoding that aim to improve its expressive capacity  \citep{lozano-cruz_practical_2026}, which we will briefly discuss.

In \textbf{basis encoding}, the simplest type of classical data representation, each component of a binary string $\mathbf{x} \in \{0,1\}^n$ is directly mapped to one of the (discrete) computational basis states of an $n$-qubit register:
\begin{equation}
	\{0,1\}^n \to \mathcal{H}_{m}\,, 
	\quad 
	\mathbf{x} \mapsto |\mathbf{x}\rangle = |x_1 x_2 \cdots x_n\rangle\,,
    \quad
    m=2^n.
\end{equation}
This approach provides an intuitive one-to-one correspondence between classical bit strings and quantum states, however, it requires that all input features be represented in binary form prior to encoding. Moreover, basis encoding consumes one qubit per binary feature, limiting scalability when dealing with high-dimensional or real-valued datasets \citep{nielsen_quantum_2010}.

\textbf{Angle encoding}, also known as rotation encoding, represents classical data by assigning each component of a real-valued vector $\mathbf{x} \in \mathbb{R}^n$ to the rotation angle of a single parameterized quantum gate, e.g. $\mathbf{R}_Y$, on each of the $n$ qubits. It typically acts on a qubit initialized in the $|0\rangle$ state, as presented in Eq. \ref{eq:angle}.
\begin{equation}
	\mathbb{R}^n \to \mathcal{H}_{m}\,, 
	\quad 
	\mathbf{x} \mapsto
	\Bigg(\bigotimes_{j=1}^{n} \mathbf{R}_Y(x_j)\Bigg)\,|0\rangle^{\otimes n}\,,
    \label{eq:angle}
\end{equation}
where $\mathbf{R}_Y(x_j)$ denotes a rotation about the y-axis of the Bloch sphere by an angle proportional to the feature value $x_j$.
This maps continuous input data directly onto the Bloch sphere, with each qubit encoding one feature as a distinct quantum rotation \citep{havlicek_supervised_2019}.

Angle encoding offers a hardware-efficient way to represent continuous data and is widely used in \acp{VQC} due to its low resource requirements.
However, since each feature requires its own qubit and rotation gate, the representational capacity is linear in the number of available qubits, limiting its scalability compared to amplitude encoding. A further limitation of angle encoding is its periodicity. For a single feature, acting on the reference state $|0\rangle$ yields
\begin{equation}
    \mathbf{R}_Y(x_j)|0\rangle
    =
    \cos\!\left(\frac{x_j}{2}\right)|0\rangle
    +
    \sin\!\left(\frac{x_j}{2}\right)|1\rangle.
    \label{eq:angle_ry}
\end{equation}
Hence, for every $k\in\mathbb{Z}$,
\begin{equation}
    \mathbf{R}_Y(x_j+2\pi k)|0\rangle
    =
    \cos\!\left(\frac{x_j}{2}+\pi k\right)|0\rangle
    +
    \sin\!\left(\frac{x_j}{2}+\pi k\right)|1\rangle
    =
    (-1)^k\mathbf{R}_Y(x_j)|0\rangle.
\end{equation}
The factor $(-1)^k$ is only a global phase and therefore physically irrelevant.

Consequently, any two inputs satisfying
\begin{equation}
    x_j' = x_j + 2\pi k_j,
    \qquad
    k_j \in \mathbb{Z},
    \quad \forall j \in \{1,\ldots,n\},
\end{equation}
are encoded into the same physical quantum state, differing only by a global phase. Equivalently, for any $\mathbf{k}\in\mathbb{Z}^n$,
\begin{equation}
    \Bigg(\bigotimes_{j=1}^{n} \mathbf{R}_Y(x_j+2\pi k_j)\Bigg)|0\rangle^{\otimes n}
    =
    (-1)^{\sum_{j=1}^{n} k_j}
    \Bigg(\bigotimes_{j=1}^{n} \mathbf{R}_Y(x_j)\Bigg)|0\rangle^{\otimes n},
    \label{eq:angle_injectivity}.
\end{equation}
so the encoding is not injective on $\mathbb{R}^n$. This implies that raw feature magnitudes can alias under angle encoding whenever they exceed the effective $2\pi$ period.
\begin{figure}[H]
    \centering
    \includegraphics[width=0.8\linewidth]{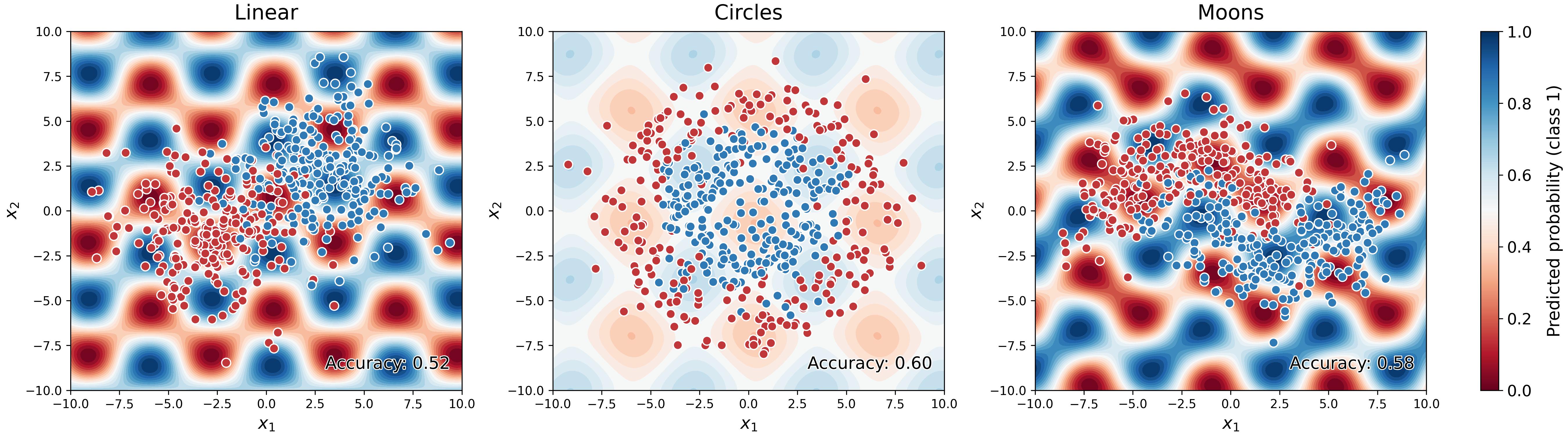}
    \caption{Visualization of the forecast of a state-vector simulated VQC with 4 qubits, 12 weights and angle encoding trained on three canonical two-dimensional binary classification problems for 20 epochs.}
    \label{fig:raw_angle_encoding}
\end{figure}
Figure \ref{fig:raw_angle_encoding} illustrates this aliasing effect on three canonical two-dimensional binary classification problems. The left panel shows a linearly separable dataset, the middle panel shows a dataset with concentric circles, and the right panel shows a dataset with interleaving half-moons.
The dots are data points in the dataset, with the fill of the dot indicating it's true class label, while the background color indicates the continuous sampled class label predictions of the \ac{VQC} after training.
For the sake of visualization, the continuous predictions are mapped to a color scale from blue (class 1) to red (class 0). The \emph{Accuracy} reported in the bottom right corner is a widely used classification metric that measures the proportion of correctly classified instances.

Without prior standardization, the raw input coordinates are mapped directly to rotation angles, so large feature values wrap repeatedly around the Bloch sphere. As a result, distant points in the original feature space can become close, or even identical, after encoding.
The learned response therefore resembles a two-dimensional sinusoidal pattern induced by the periodic input map, which aligns poorly with the target class structure and consequently exhibits weak convergence, as indicated by the Accuracy of around 0.5, equivalent to random guessing.

In practice, angle encoding therefore requires prior standardization or rescaling of the inputs so that most feature values remain in the interval $[0, 2\pi]$ \citep{wang_development_2022, sahoo_entanglement-driven_2026}. Such preprocessing does not remove the periodicity itself, but it reduces wrap-around effects and preserves meaningful variation of the encoded angles across the dataset.

\begin{figure}[H]
    \centering
    \includegraphics[width=0.8\linewidth]{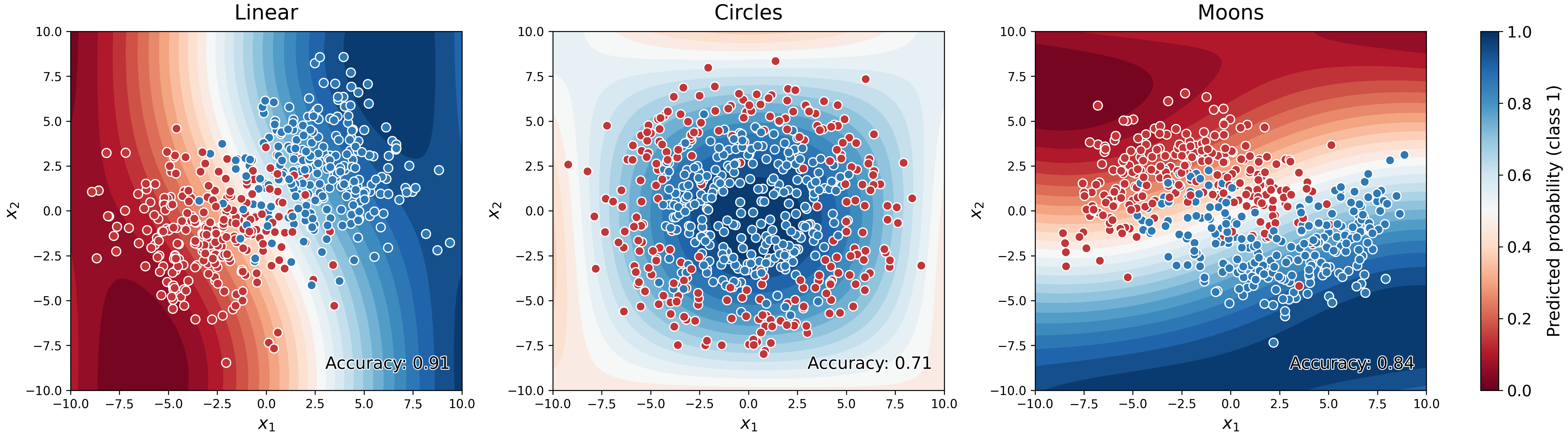}
    \caption{Visualization of the forecast of a state-vector simulated VQC with 4 qubits, 12 weights and standardized angle encoding trained on three canonical two-dimensional binary classification problems for 20 epochs.}
    \label{fig:angle_encoding}
\end{figure}

Figure \ref{fig:angle_encoding} illustrates the effect of standardization on angle encoding on the same three canonical two-dimensional binary classification problems. The angle encoding now maps the two-dimensional input features to distinct rotation angles of quantum gates, facilitating decent seperability for all three datasets, even the circles. This results in good Accuracy of around 80 \%, as indicated in the bottom right corner.
Several extensions of angle encoding have also been proposed to increase its expressive capacity by incorporating structured entangling operations into the data-embedding circuit. Examples include data reuploading \citep{perez-salinas_data_2020}, IQP encoding, which introduces feature-dependent two-qubit interactions \citep{havlicek_supervised_2019}, as well as QAOA-inspired \citep{lloyd_quantum_2020}, Ring, and Waterfall encodings \citep{venturelli_quantum_2023}, which employ different entanglement topologies to capture correlations between input features. While these approaches can construct richer feature representations than conventional angle encoding, they retain its fundamental requirement of approximately one qubit per input feature and introduce additional circuit complexity, scaling from $\mathcal{O}(LN)$ to $\mathcal{O}(LN^2)$ depending on the chosen topology and number of layers $L$.

\textbf{Amplitude encoding} is a data representation scheme that embeds the components of a classical vector into the amplitudes of a quantum state.
The idea was formalized as a state-preparation problem by \citet{mottonen_transformation_2004} and later improved on by \citet{plesch_quantum-state_2011}.

In this scheme, the components of a classical feature vector are mapped directly to the probability amplitudes of the quantum system's basis states. 
Each amplitude corresponds to one possible realization of the quantum system, so that all information about the input vector is distributed across the superposition of states rather than stored in a single register element.
Since a quantum wavefunction represents a probability distribution over all possible basis states, its squared amplitudes must collectively sum to one, i.e.\ $\sum_i |x_i|^2 = 1$.

To enforce this normalization condition for arbitrary real-valued vectors, the input is divided by its $\ell_2$-norm, ensuring that the resulting quantum state has unit total probability.

To illustrate, consider a two-qubit system: it spans a four-dimensional Hilbert space with computational basis states 
$|00\rangle$, $|01\rangle$, $|10\rangle$, and $|11\rangle$. 
Given an arbitrary real vector 
$\mathbf{x} \in \mathbb{R}^4$, 
its amplitude-encoded quantum representation is
\begin{equation*}
|\boldsymbol{\psi}(\mathbf{x})\rangle 
= \frac{1}{\|\mathbf{x}\|_{\ell_2}}\big(x_1|00\rangle + x_2|01\rangle 
+ x_3|10\rangle + x_4|11\rangle\big)\,,
\quad 
\text{where} \quad
\|\mathbf{x}\|_{\ell_2}^2 = \sum_{i=1}^{4} x_i^2~.
\end{equation*}
This means that the quantum system simultaneously encodes all entries of the classical vector as part of a coherent superposition, where the probability of measuring each basis state equals the squared magnitude of its corresponding amplitude. 
The advantage of this encoding is that a single $n$-qubit register can represent $2^n$ classical features in parallel, in this case $4$ data points with only $2$ qubits. Amplitude encoding hence provides an exponentially compact data embedding and enables efficient quantum state manipulations.

Formally, any input vector $\mathbf{x} \in \mathbb{R}^{m},~ m=2^n$ can be amplitude-encoded on an $n$-qubit quantum state as defined by
\begin{gather}
	\mathcal{A} : ~\mathbb{R}^{m}\!\setminus\!\{\mathbf{0}\} \to \mathcal{H}_{m}, 
	\quad
	\mathbf{x} \mapsto
	\frac{1}{\|\mathbf{x}\|_{\ell_2}} \sum_{i=1}^{m} x_i\,|i\rangle,
	\quad
	\|\mathbf{x}\|_{\ell_2}^2 = \sum_{i=1}^{m} x_i^2~,
    \quad
    i \in \{1,\dots,m\}.
\label{eq:amplitude_encoding}
\end{gather}
Here $\{|i\rangle\}_{i=1}^{2^{n}}$ denotes the computational basis of the $n$-qubit Hilbert space $\mathcal{H}_{2^{n}}$, with $i$ labeling the $2^{n}$ orthonormal basis states. The resulting quantum state is a point on the unit sphere in $\mathcal{H}_{2^{n}}$.

\begin{figure}[H]
    \centering
    \includegraphics[width=0.8\linewidth]{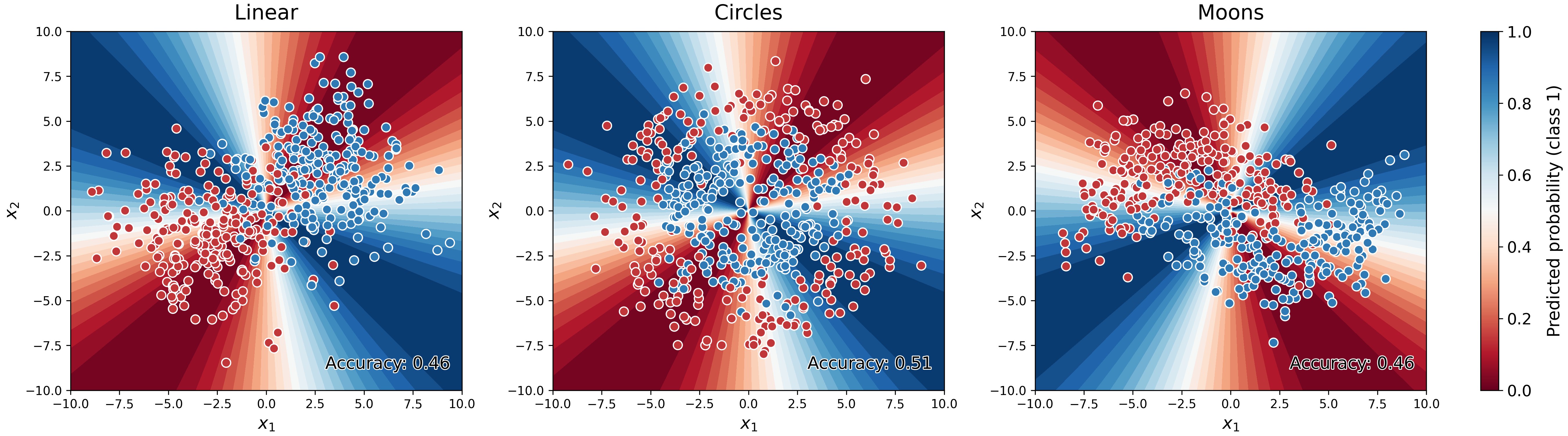}
    \caption{Visualization of the forecast of a state-vector simulated VQC with 4 qubits, 12 weights and amplitude encoding trained on three canonical two-dimensional binary classification problems for 20 epochs.}
    \label{fig:amplitude_encoding}
\end{figure}

To translate this mathematical idea into an actual quantum circuit, the state-preparation algorithm by \cite{mottonen_transformation_2004} and \cite{plesch_quantum-state_2011} is used to construct a \ac{PQC} capable of encoding any such amplitude-encoded quantum state.
The number of qubits required for this scales logarithmically with the number of features, i.e., $n = \log_2 m$, which is exponentially more efficient than angle encoding. However, a fundamental limitation of this encoding is that all information about the magnitude $\lVert \mathbf{x} \rVert_{\ell_2}$ is lost due to the normalization step. This results in invariance under scaling of the input vector:
\begin{equation}
    \mathcal{A}(\alpha \mathbf{x}) = \mathcal{A}(\mathbf{x}) \quad \forall~ \alpha >0\,.
    \label{eq:amplitude_property}
\end{equation}
Figure \ref{fig:amplitude_encoding} illustrates the practical implications of this scale invariance by visualizing the decision boundaries of a VQC with amplitude encoding on the same three canonical two-dimensional binary classification problems shown in Fig.~\ref{fig:angle_encoding}.
Amplitude encoding maps every input vector onto the unit sphere by dividing it by its Euclidean norm.
Consequently, all points lying on the same ray through the origin are represented by the same quantum state, eliminating one degree of freedom corresponding to the radial distance from the origin.
Geometrically, this corresponds to a projection of the input space onto the unit sphere.
In the two-dimensional case, the entire Euclidean plane (except for the origin) is therefore mapped onto the unit circle, with every point along a common ray collapsing to the same location on the circle.

As illustrated in Fig.~\ref{fig:amplitude_encoding}, this geometric constraint produces the characteristic star-shaped decision boundaries.
Although the orientation of these boundaries changes as the circuit parameters are optimized, their radial symmetry is preserved throughout training because it is imposed by the encoding itself rather than by the variational circuit.
Consequently, the classifier is fundamentally unable to learn decision boundaries that depend on the distance from the origin.
This results in poor class separation on all three benchmark datasets, reflected by Accuracy values close to $0.5$, i.e., equivalent to random guessing.

These observations demonstrate that, in its raw form, amplitude encoding cannot represent arbitrary decision boundaries for low-dimensional classification problems.
In the existing quantum machine learning literature, this limitation is sometimes addressed by preceding the VQC with a classical neural network that learns a feature representation before amplitude encoding \citep{bose_quantum_2025, chen_hybrid_2026}.
The classical network effectively transforms the input into a representation in which the radial symmetry introduced by normalization is no longer detrimental to the learning task.

While effective, this approach has two notable drawbacks. First, it increases the number of trainable parameters and the overall computational complexity of the model. Second, the resulting architecture is no longer fully quantum, as a substantial part of the representation learning is delegated to a classical neural network.

Ideally, one would instead seek a classical data encoding method that preserves the favorable compression of amplitude encoding while retaining the scale information discarded by the normalization.

\subsection{Recovering Scale Information in Amplitude Encoding}

Our proposed solution is to construct a fixed (i.e., non-learned) preprocessing map
\begin{equation}
    \Phi : \mathbb{R}^{d} \rightarrow \mathbb{R}^{D},
    \quad
    D>d,
\end{equation}
which precedes amplitude encoding and is chosen such that the composition
\(\mathcal{A}\circ\Phi\) distinguishes between \(\mathbf{x}\) and \(\alpha\mathbf{x}\) for every \(\alpha\neq1\). Equivalently,
\begin{equation}
    \mathcal{A}(\Phi(\mathbf{x}))
    =
    \mathcal{A}(\Phi(\mathbf{y}))
    \quad\Longrightarrow\quad
    \mathbf{x}
    =
    \mathbf{y}.
\end{equation}
Geometrically, this means that the image of \(\Phi\) contains no two distinct vectors lying on the same positive ray. Consequently, the scale ambiguity introduced by amplitude encoding is eliminated before the data are embedded into the quantum state.

Let
\begin{equation}
    \mathbf{x},\mathbf{y}\in\mathbb{R}^d,
    \qquad \text{and let} \qquad
    f:\mathbb{R}^d\rightarrow\mathbb{R}^D
\end{equation}
be a preprocessing map preceding amplitude encoding $\mathcal{A}$. Since from Eq. \ref{eq:amplitude_property}
\begin{equation}
    \mathcal{A}(\lambda\mathbf{x})
    =
    \mathcal{A}(\mathbf{x}),
    \qquad
    \forall~\lambda>0,
\end{equation}
our objective is to construct \(f\) such that
\begin{subequations}
    \begin{equation}
        f(\mathbf{x})=f(\mathbf{\lambda x})
        ~\Longrightarrow~
        \lambda=1
        ~~\text{and}~~
    \end{equation}
    \begin{equation}
        f(\mathbf{x})=f(\mathbf{y})
        ~\Longrightarrow~
        \mathbf{x}=\mathbf{y}.
    \end{equation}
\end{subequations}
A natural construction is
\begin{equation}
    f: \mathbb{R}^d \rightarrow \mathbb{R}^{2d},
    \quad
    \mathbf{x} \mapsto
    \bigl(
    \mathbf{x},
    h(\mathbf{x})
    \bigr),
\end{equation}
where \(h:\mathbb{R}^d\rightarrow\mathbb{R}^d\) denotes a nonlinear function.

However, it is important to note that nonlinearity alone does not guarantee that radial symmetry of amplitude encoding is removed. As a counterexample, consider
\begin{equation}
    h: \mathbb{R} \rightarrow \mathbb{R},\quad x \mapsto e^x
\end{equation}
\begin{subequations}
    Then
    \begin{equation}
        h(\lambda x)
        =
        \lambda h(x)
    \end{equation}
    is equivalent to
    \begin{equation}
        e^{(\lambda-1)x}
        =
        \lambda,
    \end{equation}
    which admits nontrivial solutions
    \begin{equation}
        x
        =
        \frac{\log\lambda}{\lambda-1},
        \qquad
        \lambda\neq1.
    \end{equation}
\end{subequations}

Thus, nonlinear functions may still preserve scaling ambiguities for particular inputs.

Furthermore, a feature map consisting only of a non-linear function inherits the limitation of Amplitude encoding of being undefined for inputs equal to zero (see Eq.\ \ref{eq:amplitude_encoding}). To eliminate the occurrence of such ambiguities and improve expressiveness, we instead combine two nonlinearities with distinct scaling behaviour. In particular, we propose
\begin{equation}
    \Phi(\mathbf{x})
    =
    \left(
    \mathbf{x},
    \mathbf{x}^{\circ2},
    \cos(\mathbf{x})
    \right),
    \quad
    \mathbf{x} \in \mathbb{R}^d,
    \label{eq:phi_map}
\end{equation}
where \(\mathbf{x}^{\circ2}\) denotes the element-wise square. Any scaling ambiguity would have to simultaneously satisfy$(\lambda y_i)^2=\lambda y_i^2$ and $\cos(\lambda y_i)=\lambda\cos(y_i)$ for every coordinate \(i\). This feature map hence makes nontrivial scaling solutions impossible and removes the radial symmetry of amplitude encoding, while also being well-defined for all real-valued inputs.
\begin{figure}[H]
    \centering
    \includegraphics[width=0.8\linewidth]{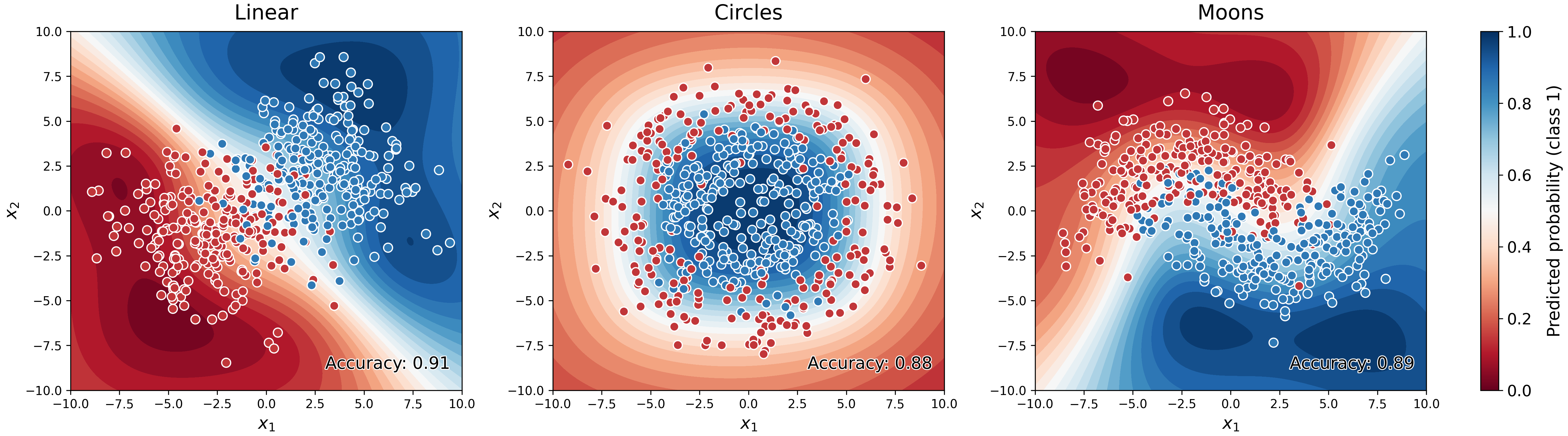}
    \caption{Visualization of the forecast of a state-vector simulated VQC with 4 qubits, 12 weights and the amplitude encoded feature map $\Phi(\mathbf{x}) = \bigl(\mathbf{x},\mathbf{x}^{\circ2},\cos(\mathbf{x})\bigr)$ trained on three canonical two-dimensional binary classification problems for 20 epochs.}
    \label{fig:phi_map}
\end{figure}
As illustrated in Figure \ref{fig:phi_map}, the proposed feature map produces excellent Accuracy values of around 90 \% and well-separated decision boundaries across all three canonical datasets, demonstrating its effectiveness in preserving scale information. On the non-linear toy datasets it even significantly surpasses angle encoding in terms of Accuracy, while still leveraging the efficiency and favourable scaling properties of amplitude encoding.

The proposed feature map is a bit less efficient than raw amplitude encoding due to the tripled dimensionality, but has the same favourable scaling charactersitics and hence requires significantly fewer qubits than angle encoding for larger datasets, while still preserving the information lost in standard amplitude encoding. This makes it a very promising approach for high-dimensional data embedding in quantum machine learning applications.

To verify that the proposed feature map not only improves computational efficiency, but also facilitates the expressive power required for complex learning tasks, the following section presents a systematic empirical evaluation on a diverse set of higher dimensional benchmark classification datasets.


\section{Design of the Empirical Validation}

\acp{VQC} yield probabilistic measurement outcomes and can therefore be used naturally for classification. We benchmark angle encoding, amplitude encoding, and amplitude encoding preceded by our proposed feature map $\Phi(\mathbf{x})$ on a range of classification tasks, namely the binary \emph{Breast Cancer} dataset, the multi-class \emph{Wine} and \emph{Digits} datasets, and a synthetic binary dataset. To isolate the effect of encoding, all quantum models share the same variational circuit architecture and optimization procedure, only the encoding strategy and associated input transformation differ. The classical reference is a feed-forward neural network, a widely used universal function approximator \citep{hornik_multilayer_1989}.

The classical reference model has two fully connected hidden layers with 16 neurons each and dropout of $0.15$.
Depending on input dimensionality and the number of output classes, it contains between 547 and 2,002 trainable parameters.
All quantum models use the same variational ansatz.
The preprocessed input is embedded through angle encoding, conventional amplitude encoding, or amplitude encoding preceded by our proposed feature map.
The ansatz consists of trainable single-qubit rotations around the $Y$ axis and CNOT entangling layers in a reverse-linear entanglement topology.
All evaluated VQCs contain exactly $32$ trainable rotation parameters.
No trainable classical layers are introduced before or after the quantum circuit.

For angle encoding, each feature requires a separate qubit.
The preprocessed input is therefore reduced by PCA to at most eight principal components.
Amplitude encoding requires fewer qubits, making PCA unnecessary.
For a feature vector of dimensionality $d$, the smallest register capable of representing all components contains $n = \left\lceil \log_2 d \right\rceil$ qubits.
The feature vector is padded to the corresponding power-of-two dimension where necessary.
$\Phi(\mathbf{x})$ may require a few additional qubits but retains logarithmic scaling.

All classical and quantum models are trained for $40$ epochs using mini-batches of $32$ samples. The VQCs are implemented in PennyLane and evaluated using GPU-accelerated state-vector simulation with the \emph{lightning.gpu} backend on an NVIDIA Quadro RTX 8000 GPU. This eliminates finite-shot uncertainty and hardware noise, allowing the comparison to focus on the encoding methods.

Hyperparameters are optimized using $20$ Optuna trials \citep{akiba_optuna_2019} on the validation set within each fold of a stratified, shuffled six-fold cross-validation, with balanced accuracy as the objective.
The selected configuration is retrained on the complete outer training fold before evaluation on the held-out test partition.
For the classical neural network, the learning rate and weight-decay coefficient are optimized over logarithmically sampled search spaces.
For the quantum models, only the learning rate is optimized.

The first three datasets were obtained from the scikit-learn collection. \emph{Breast Cancer} contains 569 samples, 30 numerical features describing cell nuclei in breast-mass images, and binary malignant/benign labels. \emph{Wine} contains 178 samples, 13 continuous chemical measurements, and three cultivar classes. \emph{Digits} contains 1,797 grayscale images with a resolution of \(8\times8\) pixels, represented by 64 pixel-intensity features, and ten digit classes.

The synthetic non-linear dataset provides a controlled binary benchmark with 2,400 samples and 12 continuous input features. Each feature was sampled independently from a uniform distribution $x_j \sim \mathcal{U}(-2.5,2.5)$.
Class labels were generated from a nonlinear score \(s(\mathbf{x})\), defined as
\begin{align}
s(\mathbf{x}) ={}&
1.2\sin(x_1)
+0.9\cos(x_2x_3)
-0.8x_4^2
+0.6e^{-|x_5|}
+0.5\tanh(x_6x_7)\\\nonumber
&-0.4\sin(x_8+x_9)
+0.3x_{10}x_{11}
-0.2x_{12}^3
+\varepsilon , \quad \varepsilon \sim \mathcal{N}(0,0.02)
\end{align}
Binary labels were obtained by thresholding the noisy latent score at its empirical median.
For all datasets, numerical features were mean-imputed and standardized to zero mean and unit variance, whereas categorical variables were imputed using their most frequent value and subsequently one-hot encoded.

Performance is quantified using accuracy, balanced accuracy, macro-averaged F1 score, precision, recall, ROC-AUC, and the area under the precision-recall curve (AUPR).
For multiclass datasets, one-vs-rest ROC-AUC and macro-averaged AUPR were used.
Reported values correspond to the mean and standard deviation across the six cross-validation folds.

Pairwise classifier differences were assessed using out-of-fold predictions and an \emph{exact McNemar test}. The six pairwise comparisons within each dataset were adjusted using the Holm--Bonferroni procedure, with adjusted \(p<0.05\) considered statistically significant.


\section{Results and Discussion}

Table~\ref{tab:results} summarizes the predictive performance of the classical neural network and the three quantum encoding approaches across the four evaluated datasets.
Among the quantum models the proposed feature map substantially improves upon conventional amplitude encoding on three of the four datasets.
Raw amplitude encoding, in turn, generally performs better than angle encoding under the computationally constrained evaluation setting considered here.

\begin{table*}[ht]
\centering
\small
\caption{Evaluation results across datasets, reported as
mean (standard deviation) over six-fold stratified cross-validation.}
\label{tab:results}
\begin{tabular}{llrrr}
\toprule
Dataset & Model & Balanced Accuracy & ROC-AUC\footnotemark[1] & AUPR\footnotemark[1] \\
\midrule
Breast Cancer
& Classical NN\footnotemark[2]              & 0.9662 (0.0214) & 0.9956 (0.0062) & 0.9971 (0.0043) \\
\cdashline{2-5}
\addlinespace[3pt]
& VQC + Angle Encoding                 & 0.6110 (0.0351) & 0.7310 (0.0296) & 0.8291 (0.0422) \\
& VQC + Amplitude Encoding             & 0.6991 (0.0255) & 0.7859 (0.0634) & 0.8437 (0.0652) \\
& VQC + Our approach              & \textbf{0.8834} (0.0639) & \textbf{0.9692} (0.0153) & \textbf{0.9800} (0.0110) \\
\midrule
Wine
& Classical NN\footnotemark[2]              & 0.9653 (0.0326) & 1.0000 (0.0000) & 1.0000 (0.0000) \\
\cdashline{2-5}
\addlinespace[3pt]
& VQC + Angle Encoding                 & 0.6816 (0.0945) & 0.8658 (0.0647) & 0.7792 (0.0963) \\
& VQC + Amplitude Encoding             & 0.7600 (0.0774) & 0.9031 (0.0412) & 0.8425 (0.0650) \\
& VQC + Our approach              & \textbf{0.9320} (0.0259) & \textbf{0.9792} (0.0067) & \textbf{0.9707} (0.0107) \\
\midrule
Digits
& Classical NN\footnotemark[2]              & 0.9408 (0.0095) & 0.9970 (0.0017) & 0.9829 (0.0083) \\
\cdashline{2-5}
\addlinespace[3pt]
& VQC + Angle Encoding                 & 0.2175 (0.0420) & 0.7130 (0.0129) & 0.2620 (0.0190) \\
& VQC + Amplitude Encoding             & 0.2751 (0.0342) & \textbf{0.7480} (0.0331) & \textbf{0.3239} (0.0185) \\
& VQC + Our approach              & \textbf{0.3100} (0.0308) & 0.7268 (0.0452) & 0.3127 (0.0319) \\
\midrule
Synthetic
& Classical NN\footnotemark[2]              & 0.8462 (0.0171) & 0.9244 (0.0153) & 0.9274 (0.0182) \\
\cdashline{2-5}
\addlinespace[3pt]
& VQC + Angle Encoding                 & 0.6458 (0.0410) & 0.7056 (0.0275) & 0.6866 (0.0238) \\
& VQC + Amplitude Encoding             & 0.5996 (0.0336) & 0.6469 (0.0446) & 0.6625 (0.0539) \\
& VQC + Our approach              & \textbf{0.7717} (0.0168) & \textbf{0.8571} (0.0120) & \textbf{0.8546} (0.0152) \\
\bottomrule
\end{tabular}
{\small \\\vspace{2pt}\footnotemark[1] For multiclass datasets (Wine and Digits), ROC-AUC denotes one-vs-rest ROC-AUC and AUPR denotes macro-averaged AUPR.\\
\footnotemark[2] Classical NN is included for reference; the best QNN result per dataset is shown in bold.}
\end{table*}

As expected, the classical neural network achieves consistently strong predictive performance. None of the quantum models consistently reaches this reference. The objective of this study, however, is not to demonstrate a quantum advantage, but to determine how much predictive capability is retained under different encoding schemes and whether the information loss of conventional amplitude encoding can be mitigated.

Among the conventional quantum encodings, amplitude encoding generally outperforms angle encoding, with the synthetic dataset being the exception.
This comparison must be interpreted in light of the different resource requirements.
Performance differences may reflect information discarded by PCA, but this constraint represents the practical trade-off under limited qubit or time resources.

The most important result is the systematic improvement obtained by combining amplitude encoding with the proposed feature map $\Phi(\mathbf{x})$.
Balanced accuracy increases from $0.6991$ to $0.8834$ on Breast Cancer, from $0.7600$ to $0.9320$ on Wine, and from $0.5996$ to $0.7717$ on Synthetic. The corresponding ROC-AUC values increase from $0.7859$ to $0.9692$, from $0.9031$ to $0.9792$, and from $0.6469$ to $0.8571$, respectively. Compared with conventional amplitude encoding, the proposed approach produces significantly more accurate out-of-fold predictions on all four datasets after Holm correction ($p_{\mathrm{Holm}}<0.05$ in each case). On Wine, our proposed QNN reaches a balanced accuracy of $0.9320$, close to the classical NN value of $0.9653$.


\section{Concluding Remarks}

This work identifies a fundamental representational limitation of amplitude encoding and introduces an efficient, non-trainable feature map to mitigate it. Although amplitude encoding represents classical feature vectors with logarithmically many qubits, normalization removes radial information by mapping all positively scaled versions of an input to the same quantum state. Our geometric and empirical analyses show that this loss can limit the expressiveness of the subsequent \acp{VQC}.

The proposed feature map recovers this information while preserving the favorable logarithmic qubit scaling of amplitude encoding. Across the evaluated benchmarks, it improved balanced accuracy over conventional amplitude encoding and substantially narrowed the gap to the classical neural-network reference on several datasets. These empirical results show that radial information remains relevant in practice and that its recovery improves predictive performance without adding additional trainable parameters.

More broadly, qubit efficiency alone is not a sufficient criterion for evaluating quantum data encodings: compression is useful only when task-relevant information is retained. The proposed approach provides a favorable compromise between compactness and representational fidelity. Although the remaining gap to classical neural networks suggests that encoding alone may not be able to fully resolve the current limitations of \acp{QNN}, the results demonstrate that careful encoding design can meaningfully improve the expressiveness of quantum models and should be considered alongside circuit architecture and optimization strategies in future work.


\subsection*{Reproducibility statement}

The source code and experimental materials required to reproduce this work are available at \url{https://anonymous.4open.science/r/quantum-encoding}. The repository includes implementations of the proposed feature map, the quantum and classical models, data preprocessing, hyperparameter optimization, cross-validation, and evaluation procedures. The Breast Cancer, Wine, and Digits datasets are obtained through scikit-learn, while the synthetic dataset can be generated using the provided code.

\clearpage


\bibliographystyle{unsrtnat}
\bibliography{QNNenc}
\urlstyle{same} 






\end{document}